\documentclass[nohyper]{JHEP3}

\usepackage{amsmath,amssymb}
\usepackage{graphicx}
\usepackage{cite}

\newcommand{\be}{\begin{equation}}
\newcommand{\ee}{\end{equation}}
\newcommand{\bea}{\begin{eqnarray}}
\newcommand{\eea}{\end{eqnarray}}
\newcommand{\epi}{\epsilon_{i}} 
\newcommand{\efi}{\epsilon_{f_i}}
\newcommand{\esi}{\epsilon_{s_i}} 

 \newcommand{\fh}{f_h^{eq}}
 \newcommand{\fht}{f_{\tilde{h}}^{eq}}
\newcommand{\fLeq}{f_{\ell}^{eq}} 
\newcommand{\fLteq}{f_{\widetilde{\ell}}^{eq}}

\title{Flavoured Soft Leptogenesis}
\author{Chee Sheng Fong 
\\
C.N. Yang Institute for Theoretical Physics\\
  State University of New York at Stony Brook\\
  Stony Brook, NY 11794-3840, USA,\\
  E-mail: \email{fong@insti.physics.sunysb.edu}}
\author{M.~C.~Gonzalez-Garcia\\
C.N. Yang Institute for Theoretical Physics\\
  State University of New York at Stony Brook\\
  Stony Brook, NY 11794-3840, USA,\\
{\rm and:}  
  Instituci\'o Catalana de Recerca i Estudis Avan\c{c}ats (ICREA), \\
  Departament d'Estructura i Constituents de la Mat\`eria,
  Universitat de Barcelona,\\
  Diagonal 647, E-08028 Barcelona, Spain\\
  E-mail: \email{concha@insti.physics.sunysb.edu}}

\abstract{We study the impact of flavour in 
``soft leptogenesis'' 
(leptogenesis induced by soft supersymmetry breaking terms). 
We address the question of how flavour effects can  
affect the region of parameters in which successful 
soft leptogenesis induced by CP violation in the right-handed sneutrino 
mixing is possible. We find that for decays which occur 
in the intermediate to strong washout regimes for all flavours, 
the produced total $B-L$ asymmetry can be up to a factor ${\cal O}(30)$
larger than the one predicted with flavour effects being neglected.
This enhancement, permits slightly larger values of the required 
lepton violating  soft bilinear term. }
\preprint{%
  YITP-SB-08-18\\}

\begin{document}

\section{Introduction}

The discovery of neutrino oscillations makes leptogenesis a very
attractive solution to the baryon asymmetry problem \cite{fy,leptoreview}.  
In the standard framework it is 
usually assumed that the tiny neutrino masses are generated via the
(type I) seesaw mechanism \cite{ss} and thus the new singlet neutral
leptons with heavy (lepton number violating) Majorana masses can
produce dynamically a lepton asymmetry through out of equilibrium
decay. Eventually, this lepton asymmetry is partially converted into a
baryon asymmetry due to fast sphaleron processes.

For a hierarchical spectrum of right-handed neutrinos,  
successful leptogenesis requires generically quite heavy singlet neutrino 
masses~\cite{di}, of order $M>2.4 (0.4)\times 10^9$~GeV for vanishing
(thermal) initial neutrino densities~\cite{di,Mbound}, 
although flavour effects \cite{flavour1,flavour2,db2,oscar} 
and/or extended scenarios \cite{db1,ma} may affect this limit.
The stability of
the hierarchy between this new scale and the electroweak one is
natural in low-energy supersymmetry, but in the supersymmetric seesaw
scenario there is some conflict between the gravitino bound on the
reheat temperature and the thermal production of right-handed
neutrinos \cite{gravi}.  This is so because in a high temperature
plasma, gravitinos are copiously produced, and their late decay could
modify the light nuclei abundances, contrary to observation. This sets
an upper bound on the reheat temperature after inflation, $T_{RH} <
10^{8-10}$ GeV, which may be too low for the right-handed neutrinos to
be thermally produced.

Once supersymmetry has been introduced, leptogenesis is induced also
in singlet sneutrino decays.  If supersymmetry is not broken, the
order of magnitude of the asymmetry and the basic mechanism are the
same as in the non-supersymmetric case. However, as shown in 
Refs.\cite{soft1,soft2,soft3}, supersymmetry-breaking terms can play an 
important role in the lepton asymmetry generated in sneutrino decays
because they induce effects which are essentially different 
from the neutrino ones.  In brief, soft supersymmetry-breaking terms
involving the singlet sneutrinos remove the mass degeneracy between
the two real sneutrino states of a single neutrino generation, and
provide new sources of lepton number and CP violation. As a
consequence, the mixing between the two sneutrino states generates a
CP asymmetry in the decay, which can be sizable for a certain range of
parameters. In particular the asymmetry is large for a right-handed
neutrino mass scale relatively low, in the range $10^{5}-10^{8}$ GeV,
well below the reheat temperature limits, what solves the cosmological
gravitino problem.  Moreover, contrary to the traditional leptogenesis
scenario, where at least two generations of right-handed neutrinos are
required to generate a CP asymmetry in neutrino/sneutrino decays, in
this new mechanism for leptogenesis the CP asymmetry in sneutrino
decays is present even if a single generation is considered.  This
scenario has been termed ``soft leptogenesis'', since the soft terms
and not flavour physics provide the necessary mass splitting and
CP-violating phase.  

In general, soft leptogenesis induced by CP violation in mixing
as discussed above has the drawback that in order to generate
enough asymmetry the lepton-violating  soft bilinear coupling has to be
unconventionally small~\cite{soft1,soft2}. 
Considering the possibility of CP violation also in decay and in the 
interference of mixing and decay of the sneutrinos~\cite{soft3}, as well 
as extended scenarios~\cite{ourinvsoft,softothers}, 
may alleviate this problem.   

In Refs.~\cite{soft1,soft2,soft3} soft leptogenesis was addressed
within the `one-flavour' approximation. This one-flavour approximation
is rigorously correct only when the interactions mediated by charged
lepton Yukawa couplings are out of equilibrium.  This is not the case
in soft leptogenesis since, as mentioned above, successful
leptogenesis in this scenario requires a relatively low right-handed
neutrino mass scale.  Thus the characteristic $T$ is such that the
rates of processes mediated by the $\tau$ and $\mu$ Yukawa couplings
are not negligible implying that the effects of lepton flavours have
to be taken into account.

The impact of flavour in thermal leptogenesis in the context of the
standard see-saw leptogenesis has been recently investigated in
much detail. \cite{flavour1,flavour2,db1,oscar,flavourothers,barbieri,PU,riottoqbefla,riottosc}.
The relevant Boltzmann Equations (BE) including flavour effects 
associated to the charged lepton Yukawa couplings 
were first introduced in Ref.~\cite{barbieri}. 
Additional flavour effects 
associated to the light-to-heavy neutrino Yukawa couplings which are
particularly relevant for the case of see-saw resonant leptogenesis were
discussed in Ref.~\cite{PU}. In Ref.~\cite{flavour1,flavour2,db1}
it was further analyzed how flavour effects can
significantly affect the result for the final baryon asymmetry.

In this work  we study the impact of flavour in soft leptogenesis.
We address the question of how flavour effects can  
affect the region of parameters in which successful 
leptogenesis induced by CP violation in the right-handed sneutrino 
mixing is possible, and in particular their impact on the required value of 
the lepton-violating soft bilinear
coupling.   
The outline of the paper is as follows.  Section~\ref{sec:unfsoft} 
revisits the  soft leptogenesis scenario with CP violation
in the mixing and we present
the relevant BE describing the production of
the lepton asymmetry in this scenario without including flavour
effects. In Sec.~\ref{sec:flasoft} we discuss the way to include
flavour-dependent processes associated with the lepton Yukawa
couplings in this scenario. Finally in Sec.~\ref{sec:results} we
present our quantitative results.

\section{Unflavoured Soft Leptogenesis}
\label{sec:unfsoft}
The supersymmetric see-saw model could be described by the superpotential:
\begin{equation}
W=\frac{1}{2}M_{ij}N_{i}N_{j}+Y_{ij}
\epsilon_{\alpha\beta}N_{i}L_{j}^{\alpha}H^{\beta},
\label{eq:superpotential}
\end{equation}
where $i,j=1,2,3$ are flavour indices and $N_{i}$, $L_{i}$, $H$
are the chiral superfields for the RH neutrinos, the left-handed (LH)
lepton doublets and the Higgs doublets with 
$\epsilon_{\alpha\beta}=-\epsilon_{\beta\alpha}$
and $\epsilon_{12}=+1$. 
The corresponding soft breaking terms involving the RH sneutrinos 
$\tilde{N_{i}}$ are given by:
\begin{equation}
\mathcal{L}_{soft}=-\tilde{m}^2_{ij}\widetilde{N}_{i}^{*}\widetilde{N}_{j}
-\left(A_{ij}Y_{ij}\epsilon_{\alpha\beta}\widetilde{N}_{i}
\tilde{\ell}_{j}^{\alpha}h^{\beta}+\frac{1}{2}B_{ij}M_{ij}
\widetilde{N}_{i}\widetilde{N}_{j}+\mbox{h.c.}\right),
\label{eq:soft_terms}
\end{equation}
where $\tilde{\ell}_{i}^{T}=\left(\tilde{\nu}_{i},\tilde{\ell}_{i}^{-}\right)$
and $h^{T}=\left(h^{+},h^{0}\right)$ are the slepton and up-type
Higgs doublets.

The Lagrangian for interaction terms involving RH sneutrinos 
$\widetilde{N}_{i}$ and RH neutrinos $N_{i}$ 
in 4-component spinors is given by:
\begin{eqnarray}
\mathcal{L}_{int} & = & -Y_{ij}\epsilon_{\alpha\beta}
\left(M_{i}\widetilde{N}_{i}^{*}\tilde{\ell}_{j}^{\alpha}h^{\beta}
+\bar{\tilde{h}}^{\beta}P_{L}\ell_{j}^{\alpha}\widetilde{N}_{i}
+\bar{\tilde{h}}^{\beta}P_{L}N_{i}\tilde{\ell}_{j}^{\alpha}
+A \widetilde{N}_{i}\tilde{\ell}_{j}^{\alpha}h^{\beta}\right)+\mbox{h.c.}
\label{eq:int_basis}\end{eqnarray}
where $\ell_{i}^{T}=\left(\nu_{i},\ell_{i}^{-}\right)$, 
$\tilde{h}^{T}=\left(\tilde{h}^{+},\tilde{h}^{0}\right)$
are the lepton and fermionic partner of $h$ and $P_{L,R}$ are 
the left or right projection operator.

The sneutrino and antisneutrino states mix with mass eigenvectors
\begin{eqnarray}
\widetilde{N}_{+i} & = &
\frac{1}{\sqrt{2}}(e^{i\Phi/2}\widetilde{N}_{i}+e^{-i\Phi/2}
\widetilde{N}_{i}^{*}),\nonumber
\\ \widetilde{N}_{-i} & = &
\frac{-i}{\sqrt{2}}(e^{i\Phi/2}
\widetilde{N}_{i}-e^{-i\Phi/2}\widetilde{N}_{i}^{*}),
\label{eq:mass_eigenstates}
\end{eqnarray}
where $\Phi\equiv\arg(BM)$ and with mass eigenvalues
\begin{eqnarray}
M_{ii\pm}^{2} & = & M_{ii}^{2}+\tilde{m}_{ii}^{2}
\pm|B_{ii}M_{ii}|.
\label{eq:mass_eigenvalues}
\end{eqnarray}

From \eqref{eq:int_basis} and \eqref{eq:mass_eigenstates}, we can
write down the Lagrangian in the mass basis as
\begin{eqnarray}
\mathcal{L}_{int}=
-\frac{Y_{ij}}{\sqrt{2}}\epsilon_{\alpha\beta}&& \left\{ 
\widetilde{N}_{+i}\left[\bar{\tilde{h}}^{\beta}P_{L}\ell_{j}^{\alpha}
+(A_{ij}+M_{i})\tilde{\ell}_{j}^{\alpha}h^{\beta}\right] \right.
\nonumber \\
&& \left.
+i\widetilde{N}_{-i}\left[\bar{\tilde{h}}^{\beta}P_{L}\ell_{_{j}}^{\alpha}
+(A_{ij}-M_{i})\tilde{\ell}_{j}^{\alpha}h^{\beta}\right]\right\}
+\bar{\tilde{h}}^{\beta}P_{L}N_{i}\tilde{\ell}_{j}^{\alpha}+\mbox{h.c.}.
\label{eq:mass_basis}
\end{eqnarray}

In what follows, we will consider a single generation of $N$ 
and $\widetilde{N}$ which we label as $1$. 
We also assume proportionality of soft trilinear terms and drop the
flavour indices for the coefficients $A$ and $B$.  
As discussed in Refs.~\cite{soft1,soft2}, in this case,  after superfield 
rotations the Lagrangians (\ref{eq:superpotential}) and (\ref{eq:soft_terms}) 
have a unique independent physical CP violating phase:
\begin{equation}
\phi={\rm arg}(A B^*)
\label{eq:CPphase}
\end{equation}
which we chose to assign to $A$.

Neglecting supersymmetry breaking effects in the right sneutrino masses
and in the vertex, the total singlet sneutrino decay width is given by  
\be
\Gamma_{\widetilde{N}_+}
=\Gamma_{\widetilde{N}_-}\equiv \Gamma_{\widetilde{N}}
=\frac {\displaystyle \sum_k |M| |Y_{1k}|^2}{\displaystyle 4 \pi} \ .
\label{eq:gamma}
\ee
\subsection{The CP asymmetry}
As discussed in Ref.\cite{soft2}, when $\Gamma \gg \Delta
M_{\pm}\equiv M_+ -M_- $, the two singlet sneutrino states are not
well-separated particles. In this case, the result for the asymmetry
depends on how the initial state is prepared. In what follows we will
assume that the sneutrinos are in a thermal bath with a thermalization
time $\Gamma^{-1}$ shorter than the typical oscillation times, $\Delta
M_\pm^{-1}$, therefore coherence is lost and it is appropriate to
compute the CP asymmetry in terms of the mass eigenstates
Eq.(\ref{eq:mass_eigenstates}).

As we will see below, the CP asymmetry produced in the decay of the 
state $\widetilde{N}_{i=\pm}$ which enters into the BE
is given by:
\be
\label{epi}
\epi = \frac{\displaystyle \sum_{a_k,k} \gamma(\widetilde{N}_i \rightarrow a_k)
- \gamma(\widetilde{N}_i \rightarrow \bar{a}_k)}
{\displaystyle \sum_{a_k,k} \gamma(\widetilde{N}_i \rightarrow a_k)
+ \gamma(\widetilde{N}_i \rightarrow \bar{a}_k)} \ , 
\ee
where $a_k\equiv s_k,f_k$ with $s_k=\tilde{\ell}_k h$ and 
$f_k=\ell_k \tilde h$ and we denote by $\gamma$ the thermal averaged rates.  
For convenience we also define the fermionic and scalar CP asymmetries 
in the decay of each 
$\widetilde{N}_i$ as
\begin{eqnarray} 
\esi & = &\frac{\displaystyle \sum_k |\hat{\mathcal M}_i
(\widetilde{N}_i \to s_k )|^2 -
|\hat{\mathcal M}_i(\widetilde{N}_i  \to  \bar s_k |^2}
{\displaystyle \sum_k|\hat{\mathcal M }_i(\widetilde{N}_i \to s_k)|^2 +
|\hat{\mathcal M}_i(\widetilde{N}_i  \to  \bar s_k)|^2}
\\ 
\efi & = &\frac{\displaystyle \sum_k |\hat{\mathcal M}_i
(\widetilde{N}_i \to f_k )|^2 -
|\hat{\mathcal M}_i(\widetilde{N}_i  \to  \bar f_k |^2}
{\displaystyle \sum_k|\hat{\mathcal M }_i(\widetilde{N}_i \to f_k)|^2 +
|\hat{\mathcal M}_i(\widetilde{N}_i  \to  \bar f_k)|^2}
 \label{asymdefhat} \ .
\end{eqnarray}
Notice that $\esi$ and $\efi$ are defined in terms of decay amplitudes,
without the phase-space factors which, as we will see, are crucial to
obtain a non-vanishing CP asymmetry~\cite{soft1,soft2}. 

Neglecting supersymmetry breaking in vertices, the total asymmetry 
$\epi$  generated in the decay of the singlet sneutrino 
$\widetilde{N}_i$ can then be written as: 
\be
\epi = \frac{\esi c_{s_i} + \efi c_{f_i}}{c_{s_i} + c_{f_i}}  \ ,
\ee
where $c_{s_i}, c_{f_i}$ are the phase-space factors of the scalar and 
fermionic channels, respectively.  

We compute the CP asymmetry following the effective field theory
approach described in \cite{pi}, which takes into account the CP
violation due to mixing of nearly degenerate states by using resumed
propagators for unstable (mass eigenstate) particles.  The decay
amplitude $\hat{\mathcal M}_i^a$ of the unstable external state
$\widetilde{N}_i$ defined in Eq.~(\ref{eq:mass_eigenstates}) into a final
state $a$ is described by a superposition of amplitudes with stable
final states: 
\be
\label{resum}
\hat{\mathcal M}_i(\widetilde{N}_i \rightarrow a) = \mathcal M_i^a 
-  \sum_{j\neq i} \mathcal M_j^a 
\frac{i \Pi_{ij}}{M_i^2 - M_j^2 + i \Pi_{jj}} \ , 
\ee
where $\mathcal M_i^a$ are the tree level decay amplitudes
and $\Pi_{ij}$ are the absorptive parts of the two-point functions
for $i,j=\pm$.
The amplitude for the decay into the conjugate final state is 
obtained from (\ref{resum}) by the replacement $\mathcal M_i^a \rightarrow 
\mathcal M_i^{a*}$.

Neglecting supersymmetry breaking in vertices 
and keeping only the lowest 
order contribution in the soft terms we find the known result
~\cite{soft1,soft2}
\begin{equation}
\epsilon_{s_+}=\epsilon_{s_-}=-\epsilon_{f_+}=-\epsilon_{f_-}
\,\equiv\, \bar\epsilon\, =\, \frac{{\rm Im}A}{M}
\frac{4\Gamma B}{4B^{2}+\Gamma^{2}}.
\label{eq:epsilon_f_T0}
\end{equation}

As long as we neglect the zero temperature lepton and slepton masses
and  small Yukawa couplings,  
the phase-space factors of the final states are flavour independent
and they are the same for $i=\pm$. 
After  including finite temperature effects in the approximation of  
decay at rest of the $\widetilde N_\pm$ they are given by:
\bea
c_{f_+}(T)=c_{f_-}(T)\equiv c_f(T)
&=&(1-x_{\ell} -x_{\tilde{h}})\lambda(1,x_{\ell},x_{\tilde{h}})
\left[ 1-\fLeq\right] \left[ 1-\fht\right] 
\label{cfeq}\\
c_{s_+}(T)=c_{s_-}(T)\equiv c_s(T)&=&\lambda(1,x_h,x_{\tilde{\ell}})
\left[ 1+\fh\right] \left[ 1+\fLteq\right]
\label{cbeq}
\eea
where
\bea
f^{eq}_{h,\tilde{\ell}}&=&\frac{1}{\exp[E_{h,\tilde{\ell}}/T]-1}
\label{eq:fHeq}\\
f^{eq}_{\tilde h,\ell}&=& \frac{1}{\exp[E_{\tilde h,\ell}/T]+1}  
\label{eq:fheq}
\eea
are the  Boltzmann-Einstein and Fermi-Dirac equilibrium distributions,
respectively, and 
\bea
&E_{\ell,\tilde h}=\frac{M}{2} (1+x_{\ell,\tilde{h}}-
x_{\tilde h,\ell}), ~~~
E_{h,\tilde{\ell}}=\frac{M}{2} (1+x_{h ,\tilde{\ell}}-
x_{\tilde{\ell},h})&\\
&\lambda(1,x,y)=\sqrt{(1+x-y)^2-4x},~~~
x_a\equiv \frac{m_a(T)^2}{M^2}&
\eea
The thermal masses for the relevant supersymmetric degrees of
freedom are \cite{thermal}:
\bea
m_h^2(T)=2 m_{\tilde h}^2(T)&=& \left(\frac{3}{8}g_2^2+\frac{1}{8}g_Y^2
+\frac{3}{4}\lambda_t^2\right) \, T^2\; ,\\
m_{\tilde{\ell}}^2(T)=2 m_\ell^2(T)&=& \left(\frac{3}{8}g_2^2+\frac{1}{8}g_Y^2
\right)\, T^2\; .
\eea
Here $g_2$ and $g_Y$ are gauge couplings and $\lambda_t$ is the top Yukawa,
renormalized at the appropriate high-energy scale. 

As we will see in Sec.~\ref{subsec:beunf} the contribution 
to the relevant BE for the lepton number scalar and fermion 
asymmetries  can be factorized respectively as:
\be 
\epsilon_{s}(T) \equiv
\frac {\displaystyle\sum_k \gamma(\widetilde{N}_\pm
\rightarrow s_k) - \gamma(\widetilde{N}_\pm \rightarrow \bar{s}_k)}
{\displaystyle \sum_{a_k,k} \gamma(\widetilde{N}_\pm \rightarrow a_k) +
\gamma(\widetilde{N}_\pm \rightarrow \bar{a}_k)}
\equiv\, \bar\epsilon\,\frac{c_{s}(T)}{c_{s}(T)+c_{f}(T)} 
\ee
and 
\be
\epsilon_{f}(T) \equiv
\frac {\displaystyle \sum_k\gamma(\widetilde{N}_\pm
\rightarrow f) - \gamma(\widetilde{N}_\pm \rightarrow \bar{f})}
{\displaystyle \sum_{a_k,k} \gamma(\widetilde{N}_\pm \rightarrow a_k) +
\gamma(\widetilde{N}_\pm \rightarrow \bar{a})} 
\equiv \,-\bar\epsilon\, \frac{c_{f}(T)}{c_{s}(T)+c_{f}(T)}  \, .
\ee
The total CP asymmetry generated in the decay of any of the sneutrino 
$\widetilde{N}_{\pm}$ is then:
\be
\epsilon (T) = \,\bar \epsilon\, \ \frac{c_{s}(T) - c_{f}(T)}{c_{s}(T) + c_{f}(T)} \equiv \bar\epsilon\; \Delta_{BF} (T)\ .
\label{eq:asymunf}
\ee

In this derivation we have neglected thermal corrections to the CP 
asymmetry from the loops, 
i.e., we have computed the imaginary part of the one-loop graphs using 
Cutkosky cutting rules at $T=0$. These corrections 
are the same for scalar and fermionic decay channels,
since only bosonic loops contribute to the wave-function diagrams
in both cases, so they are not expected to introduce significant changes.

\subsection{The Boltzmann Equations}
\label{subsec:beunf}
We next write the relevant classical BE describing the decay, inverse 
decay and scattering processes involving the sneutrino states.  

As mentioned above we assume that the sneutrinos are in a thermal bath with a 
thermalization time shorter than the oscillation time. Under this assumption
the initial states can be taken as being the mass eigenstates in 
Eq.~(\ref{eq:mass_eigenstates}) and we write the corresponding equations for
those states and the scalar and fermion lepton numbers. 
The $CP$ fermionic and scalar asymmetries for each $\widetilde{N}_i$ defined
at $T=0$ are those given in Eq.~(\ref{eq:epsilon_f_T0}).

The BE describing the evolution of the number density 
of particles in the plasma are:
\begin{eqnarray}
  \frac{dn_X}{dt} + 3Hn_X & = & \sum_{j,l,m} 
\Lambda^{X j \dots}_{l m \dots} \left[ f_l f_m \dots (1\pm f_X) (1\pm f_j)
    \dots W(l m \dots \to X j \dots)   - \right. \nonumber \\
 & - & \left.   f_X f_j \dots (1\pm
  f_l)(1\pm f_m) \dots W(X j \dots \to l m \dots) \right] \nonumber
\end{eqnarray}
where, 
\bea
\Lambda^{X j \dots}_{l m \dots} & = & 
\int \frac{d^3 p_X}{(2\pi)^3 2E_X} 
\int \frac{d^3 p_j}{(2\pi)^3 2E_j} \dots 
\int \frac{d^3 p_l}{(2\pi)^3 2E_l}
\int \frac{d^3 p_m}{(2\pi)^3 2E_m} \dots  \; ,\nonumber
\eea
and $W(l m \dots \to X j \dots)$ is the squared transition amplitude summed
over initial and final spins. 
In what follows we will use the notation of Ref.\cite{kolb}. 
We we will assume that the Higgs and higgsino fields are in 
thermal equilibrium with distributions given in 
Eqs.~(\ref{eq:fHeq}) and ~(\ref{eq:fheq}) respectively. 
Strictly speaking this implies that we are not including all the effects 
associated with spectator processes~\cite{spec1,spec2}. 
For the leptons and sleptons we assume that they are in kinetic
equilibrium and we account for their asymmetries by introducing a chemical 
potential for the leptons, $\mu_\ell$, and sleptons, $\mu_{\tilde \ell}$:
\begin{eqnarray}
f_{\ell}  =  \frac{1}{e^{(E_{\ell}-\mu_{\ell})/T}+1}, &\;\;\;\;\;\;\;\; & 
f_{\tilde{\ell}} =  \frac{1}{e^{(E_{\tilde{\ell}}
-\mu_{\tilde{\ell}})/T}-1},
\end{eqnarray}
and the corresponding ones for the antiparticles with the exchange
$\mu_\ell\rightarrow -\mu_\ell$ and 
$\mu_{\tilde \ell}\rightarrow -\mu_{\tilde \ell} $ respectively.
Furthermore in order to eliminate the dependence in the expansion of the
Universe we write the equations in terms of the abundances 
$Y_X$, where $Y_X=n_X/s$. Also for convenience we use the variable
$z=M/T$.

We are interested in the evolution of sneutrinos
$Y_{\widetilde{N}_i}$, and the fermionic $Y_{L_f}$ and scalar $Y_{L_s}$
lepton numbers, defined as $Y_{L_f}=(Y_\ell-Y_{\bar\ell})$,
$Y_{L_{s}}=(Y_{\tilde\ell}-Y_{\tilde\ell^*})$. Moreover,  
in order to account for all the $\Delta L=1$
terms we also need to consider the evolution of the right-handed
neutrino $Y_N$. 

Neglecting supersymmetry breaking effects in the right sneutrino masses
and in the vertices, all the amplitudes for $N_+$ and $N_-$ are equal 
as well as their corresponding equilibrium number densities, 
$f_{\widetilde{N}_{+}}^{eq}=
f_{\widetilde{N}_{-}}^{eq}\equiv f_{\widetilde{N}}^{eq}$.
So we can define a unique BE for 
$Y_{\widetilde{N}_{\mbox{tot}}}\equiv 
Y_{\widetilde{N}_{+}}+Y_{\widetilde{N}_{-}}$. 
Thus, in total, in this unflavour case,
we have a set of four BE. 
 
The derivation of the factorization of the relevant CP asymmetries 
including the thermal effects is somehow lengthy but straight forward.
In particular one has to use that at ${\cal O}(\epsilon)$ we can neglect 
the difference between  $f_{\widetilde{N}_{\pm}}$  and 
$f^{eq}_{\widetilde{N}_{\pm}}$ in the definitions of the thermal 
average widths (see for example Ref.~\cite{ourinvsoft}). 
Many of the terms in the equations are equivalent to the ones given 
for example in  Ref.~\cite{plumacher}.~\footnote{However, some care has 
to be taken as the Eqs. in Ref.~\cite{plumacher} are given in the weak 
basis for the $\widetilde {N}$ while we give here the corresponding equations
in the mass basis.}  

Altogether we find :
\begin{eqnarray}
sHz\frac{dY_{N}}{dz} & = & 
-\left(\frac{Y_{N}}{Y_{N}^{eq}}-1\right)
\left(\gamma_{N}+4\gamma_{t}^{(0)}+4\gamma_{t}^{(1)}
+4\gamma_{t}^{(2)}+2\gamma_{t}^{(3)}+4\gamma_{t}^{(4)}\right),
\label{eq:BEN}\\
sHz\frac{dY_{\widetilde{N}_{\mbox{tot}}}}{dz} & = & 
-\left(\frac{Y_{\widetilde{N}_{\mbox{tot}}}}{Y_{\widetilde{N}}^{eq}}
-2\right)\left(\gamma_{\widetilde{N}}+\gamma_{\widetilde{N}}^{(3)}
+3\gamma_{22}+2\gamma_{t}^{(5)}+2\gamma_{t}^{(6)}
+2\gamma_{t}^{(7)}+\gamma_{t}^{(8)}+2\gamma_{t}^{(9)}\right)
\nonumber \\
&&
-\gamma_{\widetilde{N}} \frac{Y_{L_{f}}\,\epsilon_f(T)\,+Y_{L_{s}}\,\epsilon_s(T)}
{Y_{c}^{eq}}, 
\label{eq:BENt} \\
sHz\frac{dY_{L_{f}}}{dz} & = & \gamma_{\widetilde{N}}
\left[\epsilon_{f}(T)\,\left(\frac{Y_{\widetilde{N}_{\mbox{tot}}}}
{Y_{\widetilde{N}}^{eq}}-2\right)-\frac{Y_{L_{f}}}{Y_{c}^{eq}}
\frac{\gamma^f_{\widetilde{N}}}{\gamma_{\widetilde{N}}} 
\right]
\nonumber \\
 &  & -\frac{Y_{L_{f}}}{Y_{c}^{eq}}\left(\frac{1}{4}\gamma_{N}
+\frac{Y_{\widetilde{N}_{\mbox{tot}}}}{Y_{\widetilde{N}}^{eq}}
\gamma_{t}^{(5)}+2\gamma_{t}^{(6)}+2\gamma_{t}^{(7)}+
\frac{Y_{N}}{Y_{N}^{eq}}\gamma_{t}^{(3)}+2\gamma_{t}^{(4)}\right)
\nonumber \\
&&+\frac{Y_{L_{f}}-Y_{L_{s}}}{Y_{c}^{eq}}\gamma_{\mbox{MSSM}} ,
\label{eq:BELf}
\end{eqnarray}
\begin{eqnarray}
sHz\frac{dY_{L_{s}}}{dz} & = & \gamma_{\widetilde{N}}
\left[\epsilon_{s}(T)\,\left(\frac{Y_{\widetilde{N}_{\mbox{tot}}}}
{Y_{\widetilde{N}}^{eq}}-2\right)-\frac{Y_{L_{s}}}{Y_{c}^{eq}}
\frac{\gamma^s_{\widetilde{N}}}{\gamma_{\widetilde{N}}} 
\right] 
\nonumber \\
 &  & -\frac{Y_{L_{s}}}{Y_{c}^{eq}}\left(\frac{1}{4}\gamma_{N}
+\gamma_{\widetilde{N}}^{(3)}+
\frac{1}{2}
\frac{Y_{\widetilde{N}_{\mbox{tot}}}}
{Y_{\widetilde{N}}^{eq}}
\gamma_{t}^{(8)}+2\gamma_{t}^{(9)}
+2\frac{Y_{N}}
{Y_{N}^{eq}}\gamma_{t}^{(0)}+2\gamma_{t}^{(1)}+2\gamma_{t}^{(2)}\right)
\nonumber \\
 &  & -\frac{Y_{L_{s}}}{Y_{c}^{eq}}\left(2 +\frac{1}{2}
\frac{Y_{\widetilde{N}_{\mbox{tot}}}}{Y_{\widetilde{N}}^{eq}}\right)
\gamma_{22} -
\frac{Y_{L_{f}}-Y_{L_{s}}}{Y_{c}^{eq}}
\gamma_{\mbox{MSSM}}
\label{eq:BELs}
\end{eqnarray}
In the equations above $Y_{c}^{eq}\equiv\frac{15}{4\pi^{2}g_{s}^{*}}$  
and $Y^{\rm eq}_{\tilde N}(T\gg M) = 90 \zeta(3)/(4\pi^4g_s^*)$,    
where $g_{s}^{*}$ is the total number of entropic degrees of 
freedom, $g_{s}^{*}=228.75$ in the MSSM.

The different $\gamma$'s are the thermal widths for the following processes:
\begin{eqnarray}
&&\gamma_{\widetilde N}=\gamma^f_{\widetilde N}+\gamma^s_{\widetilde N}=
\gamma(\widetilde{N}_{\pm}\leftrightarrow
\bar{\tilde{h}}\ell)
+\gamma(\widetilde{N}_{\pm} \leftrightarrow h\tilde{\ell}) ,
\nonumber \\
&&\gamma^{(3)}_{\widetilde N}=\gamma( 
\widetilde{N}_{\pm}\leftrightarrow 
\tilde{\ell}^{*}\tilde{u}\tilde{q})\; , 
\nonumber \\
&&\gamma_{22} =  \gamma(\widetilde{N}_{\pm}
\tilde{\ell}\leftrightarrow\tilde{u}\tilde{q}
)=\gamma(\widetilde{N}_{\pm}
\tilde{q}^{*}\leftrightarrow\tilde{\ell}^{*}\tilde{u}
)=\gamma(\widetilde{N}_{\pm}\tilde{u}^{*}\leftrightarrow\tilde{\ell}^{*}\tilde{q}) ,
\nonumber \\
&&\gamma_t^{(5)}=\gamma(\widetilde{N}_{\pm}
\ell\leftrightarrow q\tilde{u})=\gamma(
\widetilde{N}_{\pm}\ell\leftrightarrow\tilde{q}\bar{u})\; ,
\nonumber \\
&&\gamma_t^{(6)}=\gamma(
\widetilde{N}_{\pm}\tilde{u}\leftrightarrow\bar{\ell}q)=\gamma( 
\widetilde{N}_{\pm}\tilde{q}^{*}\leftrightarrow\bar{\ell}\bar{u})\;,
\nonumber \\
&&\gamma_t^{(7)}=\gamma(
\widetilde{N}_{\pm}\bar{q}\leftrightarrow\bar{\ell}\tilde{u})=\gamma( 
\widetilde{N}_{\pm}u\leftrightarrow\bar{\ell}\tilde{q}) , 
\nonumber \\
&&\gamma_t^{(8)}=\gamma(
\widetilde{N}_{\pm}\tilde{\ell}^{*}\leftrightarrow\bar{q}u) ,
\nonumber \\
&&\gamma_t^{(9)}=\gamma(
\widetilde{N}_{\pm}q\leftrightarrow\tilde{\ell}u)= 
\gamma(\widetilde{N}_{\pm}\bar{u}\leftrightarrow\tilde{\ell}\bar{q}) ,
\nonumber \\
&&\gamma_N=\gamma(N\leftrightarrow \ell h)+
\gamma(N\leftrightarrow \tilde{\ell}^* \tilde h) ,
\nonumber \\
&&\gamma_t^{(0)}=\gamma(N\tilde{\ell}\leftrightarrow q\tilde{u})=
\gamma(N\tilde{\ell}\leftrightarrow\tilde{q}\bar{u}) ,
\nonumber \\
&&\gamma_t^{(1)}=\gamma(N\bar{q}\leftrightarrow\tilde{\ell}^{*}\tilde{u})=
\gamma(N\leftrightarrow\tilde{\ell}^{*}\tilde{q})\; ,
\nonumber \\
&&\gamma_t^{(2)}=\gamma(N\tilde{u}^{*}\leftrightarrow\tilde{\ell}^{*}q)=
\gamma(N\tilde{q}^{*}\leftrightarrow\tilde{\ell}^{*}\bar{u})\; , 
\nonumber \\
&&\gamma_t^{(3)}=\gamma(N\ell\leftrightarrow q\bar{u})\; ,
\nonumber \\
&&\gamma_t^{(4)}=\gamma (N\leftrightarrow\bar{\ell}q)=
\gamma(N\bar{q}\leftrightarrow\bar{\ell}\bar{u})\; ,
\label{eq:gammas}
\end{eqnarray}
where in all cases a sum over the CP conjugate final states is implicit.

We have included in Eqs.(~\ref{eq:BEN}--\ref{eq:BELs})  
the $\widetilde{N}_{\pm}$ and $N$ decay and 
inverse decay processes as well as all the $\Delta L=1$ scattering 
processes induced by the $top$ Yukawa coupling.
We ignore $\Delta L=1$ scattering involving gauge bosons. 
We have accounted for the dominant CP asymmetry in the mixing as generated by
the thermal effects in the $\widetilde{N}_{\pm}$  two body decays but 
we have not included the possible CP violating effects
induced by mixing in its three body decays or in its scattering
processes. $\Delta L=2$ processes involving the on-shell  
exchange of $N$ or $\widetilde{N}_{\pm}$  are already accounted for by 
the decay and inverse decay processes. The $\Delta L=2$ 
off-shell scattering processes involving the pole-subtracted s-channel 
and the u and t-channel, as well as the  
the $L$ conserving processes from $N$ and $\widetilde{N}$ pair creation 
and annihilation have not been included. The
reaction rates for these processes are quartic in the Yukawa
couplings, ie they involve factors $(YY^\dagger)^2$, and therefore can
be safely neglected as long as the Yukawa couplings are much smaller
than one, as it is the case.

The explicit expressions for 
the $\gamma$'s in Eq.~(\ref{eq:gammas}) can be found, for example, 
in \cite{plumacher} for the case of 
Boltzmann-Maxwell distribution functions and neglecting Pauli-blocking 
and stimulated emission as well as the relative motion of the particles
with respect to the plasma~\footnote{
Neglecting supersymmetry breaking effects in the right sneutrino masses
and in the vertices, it can be shown that the thermal widths for 
the sneutrino mass eigenstates and weak eigenstates are the same}.
With these approximations, for example:
\begin{eqnarray}
\gamma_{\widetilde{N}}  =  
n_{\widetilde{N}}^{eq}\, 
\Gamma_{\widetilde{N}}
\, \frac{\mathcal{K}_{1}(z)}{\mathcal{K}_{2}(z)},
&\;\;\;\;\;\;\;\;& \gamma_{N}  =  n_{{N}}^{eq}\, 
\Gamma_{{N}}
\, \frac{\mathcal{K}_{1}(z)}{\mathcal{K}_{2}(z)},
\end{eqnarray}
where $\mathcal{K}_{1,2}(z)$ are the modified Bessel function of 
the second kind of order 1 and 2 and 
$\Gamma_N=\Gamma_{\widetilde{N}}$ are the zero temperature widths 
Eq.~(\ref{eq:gamma}). In our calculation we keep the thermal
masses and statistical factors on the CP asymmetries but we  
neglect them in the rest of the thermal widths,  
with  the exception of the Higgs mass the 
in the $\Delta L=1$ processes involving  a Higgs boson exchange in 
the $t$-channel.
 
$\gamma_{MSSM}$ represent processes which transform leptons into scalar
leptons and vice versa (for example $[e+e\leftrightarrow\tilde{e}+\tilde{e}]$).
The rates for these reactions are larger than the ones in Eq.~(\ref{eq:gammas})
because they do not involve the Yukawa couplings $Y_{ij}$. Consequently 
they  enforce  that $Y_{L_{f}}\approx Y_{L_{s}}$.

For  $Y_{L_{f}}=Y_{L_{s}}$ we can combine the BE for 
$Y_{L_{f}}$ and $Y_{L_{s}}$ by defining
\begin{eqnarray}
Y_{L_{\mbox{tot}}} & \equiv & Y_{L_{f}}+Y_{L_{s}},
\end{eqnarray}
which obeys the BE:
\begin{eqnarray}
sHz\frac{dY_{L_{\mbox{tot}}}}{dz} & = 
& \left[\,\epsilon(T)\,\left(\frac{Y_{\widetilde{N}_{\mbox{tot}}}}
{Y_{\widetilde{N}}^{eq}}-2\right)-\frac{Y_{L_{\mbox{tot}}}}
{2Y_{c}^{eq}}\right]\gamma_{\widetilde{N}}
\nonumber \\
 &  & -\frac{Y_{L_{\mbox{tot}}}}{2Y_{c}^{eq}}
\left(\frac{1}{4}\gamma_{N}+\frac{Y_{\widetilde{N}_{\mbox{tot}}}}
{Y_{\widetilde{N}}^{eq}}\gamma_{t}^{(5)}+2\gamma_{t}^{(6)}
+2\gamma_{t}^{(7)}+\frac{Y_{N}}{Y_{N}^{eq}}\gamma_{t}^{(3)}
+2\gamma_{t}^{(4)}\right)\nonumber \\
 &  & -\frac{Y_{L_{\mbox{tot}}}}{2Y_{c}^{eq}}
\left(\frac{1}{4}\gamma_{N}+\gamma_{\widetilde{N}}^{(3)}
+\frac{1}{2}
\frac{Y_{\widetilde{N}_{\mbox{tot}}}}
{Y_{\widetilde{N}}^{eq}}
\gamma_{t}^{(8)}+2\gamma_{t}^{(9)}
+2\frac{Y_{N}}{Y_{N}^{eq}}\gamma_{t}^{(0)}+2\gamma_{t}^{(1)}
+2\gamma_{t}^{(2)}\right)\nonumber \\
 &  & -\frac{Y_{L_{\mbox{tot}}}}{2Y_{c}^{eq}}\left(2+\frac{1}{2}
\frac{Y_{\widetilde{N}_{\mbox{tot}}}}{Y_{\widetilde{N}}^{eq}}\right)
\gamma_{22}.
\label{eq:BE_L_tot}
\end{eqnarray}
Also the second line in Eq.~(\ref{eq:BENt}) can be written as 
$-\gamma_{\widetilde{N}}\epsilon(T) \frac{Y_{L_{\mbox{tot}}}}{2Y_{c}^{eq}}$. 
So in total we are left with three BE for  
$Y_N$, $Y_{\widetilde{N}_{\mbox{tot}}}$, and $Y_{L_{\mbox{tot}}}$.

The final amount of ${B}-{L}$ asymmetry generated by the decay of the 
singlet sneutrino states assuming no pre-existing
asymmetry can be parameterized as:
\begin{equation}
Y_{B-L}(z\rightarrow \infty)=-Y_{L_{\rm tot}}(z\rightarrow \infty)=
-2 \eta\, \bar\epsilon \, Y^{\rm eq}_{\tilde N}(T>>M) 
\label{eq:yb-l}
\end{equation}
where $\bar\epsilon$ is given in Eq.(\ref{eq:epsilon_f_T0})~\footnote{
The factor 2 in Eq.~(\ref{eq:yb-l}) arises from the fact that there are two 
right-handed sneutrino states while we have defined 
$Y^{\rm eq}_{\tilde N}$ for one degree of freedom. Defined this way,
$\eta$ has the standard normalizaiton $\eta\rightarrow 1$ for perfect
out of equilibrium decay.}. 

$\eta$ is a dilution factor which
takes into account the possible inefficiency in the production of the
singlet sneutrinos, the erasure of the generated asymmetry by
$L$-violating scattering processes and the temperature dependence of the 
CP asymmetry  and it is obtained by solving the array
of BE above.  Within our approximations for the 
thermal widths, $\eta$ depends on the values of the Yukawa couplings 
$(YY^\dagger)_{11}$ and the  heavy mass $M$, with the dominant dependence 
arising in the combination
\begin{equation}
(YY^\dagger)_{11}\,  v_u^2 \equiv m_{eff}\, M
\label{eq:meff}
\end{equation}
where $v_u$ is the vacuum expectation 
value of the up-type Higgs doublet, $v_u=v\, \sin\beta $ ($v$=174 GeV) .
There is a residual dependence on $M$ due to the running of 
the top Yukawa coupling as well as the thermal effects included in 
$\Delta_{BF}$ although it is very mild. 

In Fig.~\ref{fig:etaunf} we plot $|\eta|$ as a function of $m_{eff}$ for
$M=10^7$ GeV. Following Ref.~\cite{soft1,thermal}
we consider two different initial conditions for the sneutrino abundance.
In one case, one assumes that the ${\tilde N}$ population is created
by their Yukawa interactions with the thermal
plasma, and set $Y_{\tilde N}(z\rightarrow 0)=0$. The other case corresponds to
an initial $\tilde N$ abundance equal to the thermal one, 
$Y_{\tilde N}(z\rightarrow 0)= Y_{\tilde N}^{eq}(z\to 0)$~\footnote{
The (in)dependence of the final asymmetry on the exact preparation of
the initial state has been further explored in 
Ref.~\cite{BahatTreidel:2007ic}.}. 

Our results show good agreement with those in
Refs.~\cite{soft1,soft2,soft3}. 
In particular we reproduce that for zero
initial conditions, $\eta$ can take both signs depending on the value
of $m_{eff}$, thus it is possible to generate the right
sign asymmetry with either sign of ${\rm Im} A $.  For thermal initial
conditions, on the contrary, $\eta>0$ and the right asymmetry can only
be generated for ${\rm Im} A>0 $.  The plot is shown for
$\tan\beta=30$. But as long as $\tan\beta$ is not very close to one,
the dominant dependence on $\tan\beta$ arises via $v_u$ as given in
Eq.~(\ref{eq:meff}) and it is therefore very mild. For
$\tan\beta\sim {\cal O}(1)$ there is also an additional (very weak)
dependence due to the associated change in the top Yukawa coupling.

\FIGURE[t]{
\includegraphics[width=0.6\textwidth]{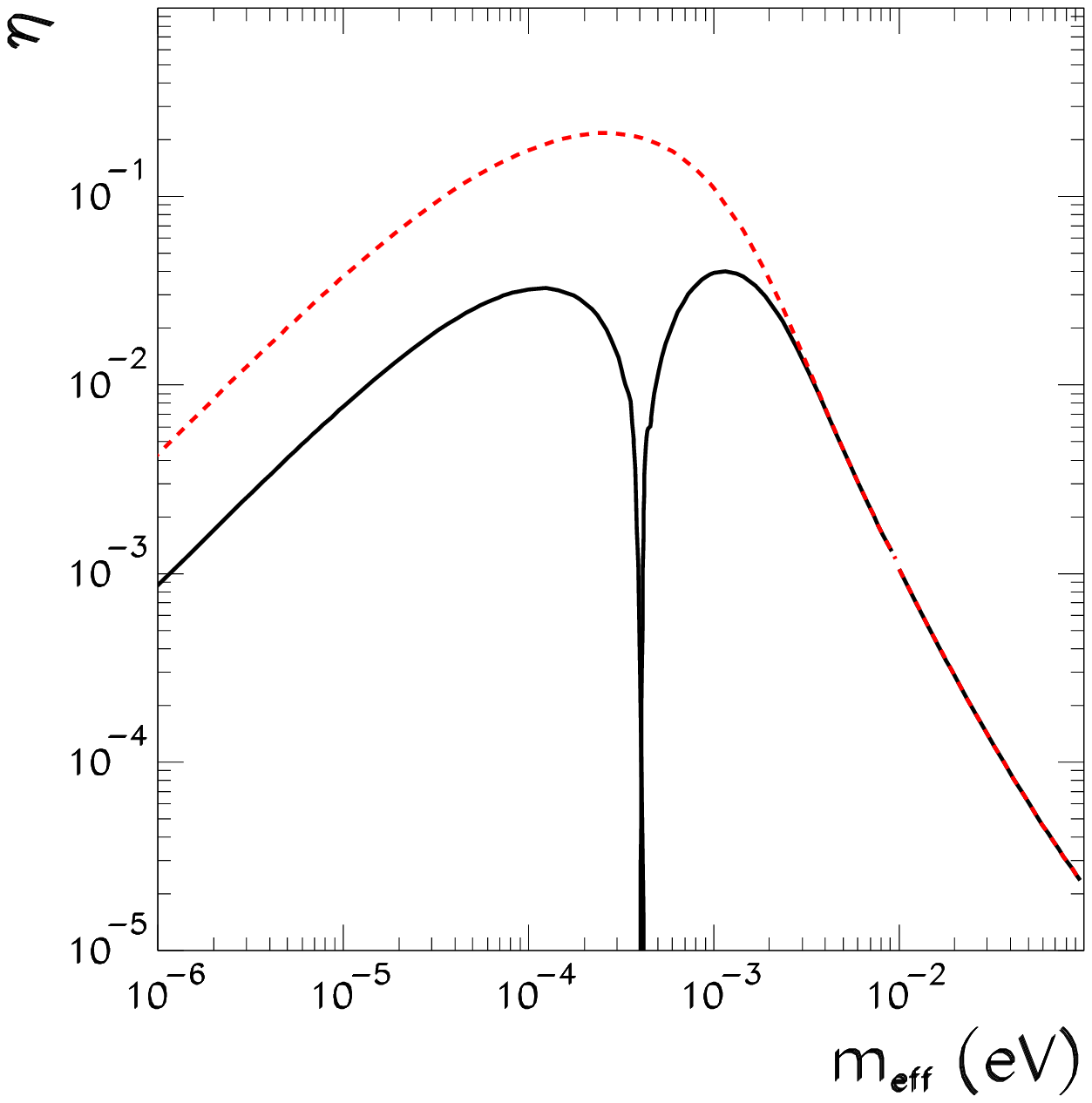}
\caption{Efficiency factor $|\eta |$ as a function of $m_{eff}$ 
for $M=10^{7}$ GeV and $\tan\beta=30$. 
The two curves correspond to  
vanishing initial $\tilde N$ abundance 
(solid black curve) and thermal initial $\tilde N$ abundance, 
(dashed red curve). 
\label{fig:etaunf}}}

After conversion by the sphaleron transitions, the final baryon asymmetry
is related to the ${B}-{ L}$ asymmetry by
\begin{equation}
Y_{B}= \frac{24+4n_H}{66+13n_H}
Y_{B-L}(z\rightarrow \infty)
=\frac{8}{23} \, Y_{B-L}(z\rightarrow \infty)
\end{equation}
where $n_H$ is the number of Higgs doublets, which is taken to be
$n_H=2$ for the MSSM in the second equality.

This has to be compared with the WMAP measurements that in
the $\Lambda$CDM model imply~\cite{lastwmap}: 
\begin{equation}
{Y_B}= (8.78\pm0.24)\times 10^{-11}
\label{eq:wmapeta}
\end{equation}

\FIGURE[t]{
\includegraphics[width=\textwidth]{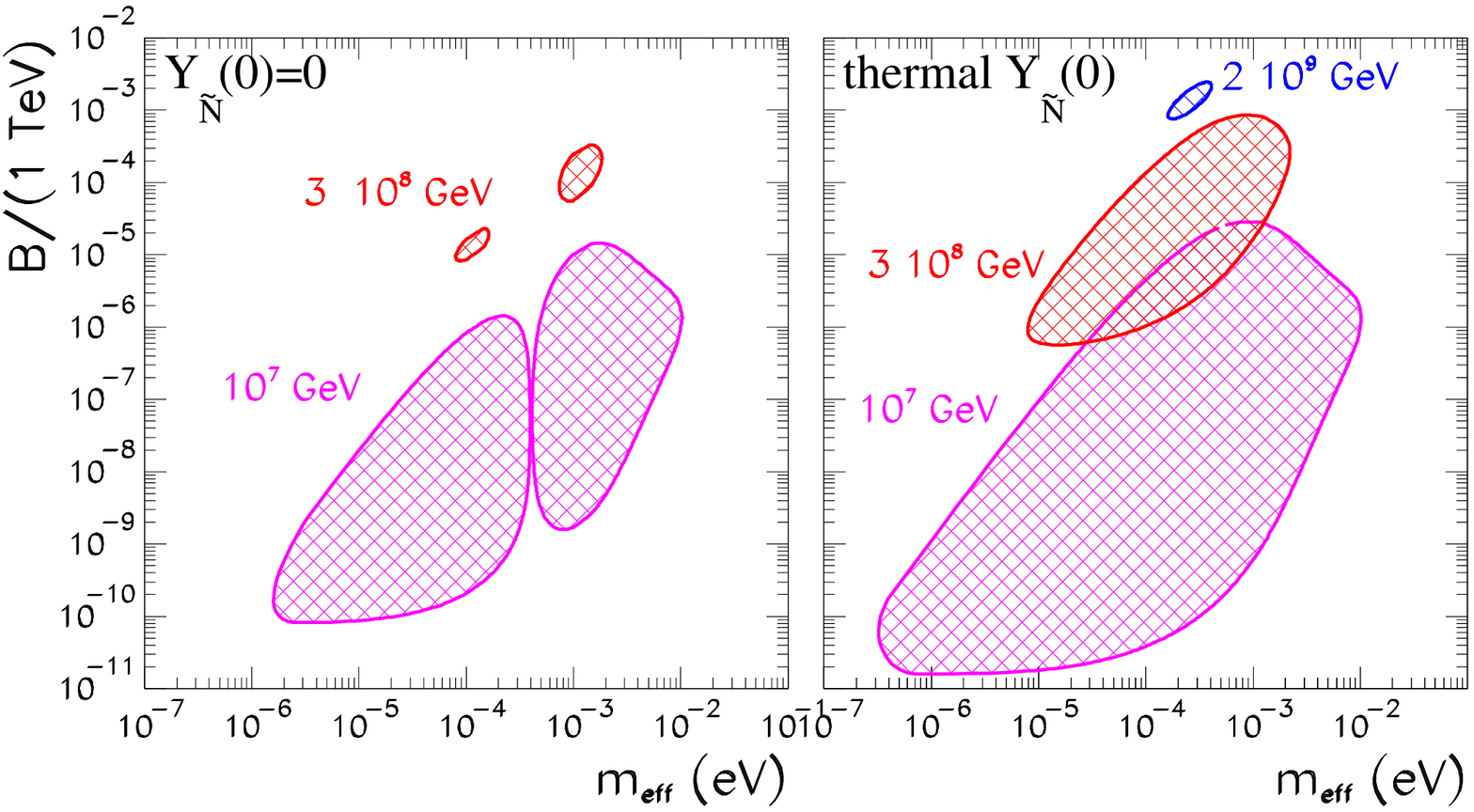}
\caption{$B, m_{eff}$ regions in which successful soft leptogenesis can
be achieved. We take $|{\rm Im} A|=10^3$ GeV and $\tan\beta=30$ and different
values of $M$  as labeled in the figure.
The two panels correspond to  vanishing initial $\tilde N$ abundance 
(left) and thermal initial $\tilde N$ abundance, 
(right). \label{fig:Bmeffunf}}}

We plot in  Fig. \ref{fig:Bmeffunf} the range of parameters
$B$ and $m_{eff}$ for which enough asymmetry is
generated, $Y_B\geq 8.54\times 10^{-11}$.
We show the ranges for several values of $M$  and for the characteristic 
value of  $|{\rm Im} A|=1$ TeV.

The figure illustrates our quantification of the known result that 
independently of the $\tilde N$ initial distributions, 
successful soft leptogenesis requires 
$M\lesssim 10^{9}$ GeV as well as $B\ll A$. Next we turn to 
the effect of flavour on these conclusions.

Before doing so, let us comment that, as pointed out in
Refs.~\cite{soft1,soft2,soft3}, these results indicate that in soft
leptogenesis, the CP asymmetry is maximal when the parameter lie on
the resonant condition $\Gamma=2 |B|$. In this case, the asymmetry is
generated by the decays of two nearly mass-degenerate $\widetilde
N$. 
It has been recently discussed in Refs.~\cite{riottoqbe} that for
resonant scenarios, the use of quantum BE~\cite{buchqbe,riottoqbe}
(QBE)  may be relevant. In particular it has been shown 
that for standard see-saw resonant leptogenesis there are differences
with the classical treament in the weak washout regime.
There, is however, no study on the literature of the impact of the use
of QBE for the case of soft leptogenesis and to discuss those is
beyond the scope of this paper. 
Thus in our  work here we study the impact of flavour in soft leptogenesis 
in the context of the classical BE as described above.
  
\section{Flavour Effects}
\label{sec:flasoft}
In the previous discussion flavour effects have been neglected.
This is only justified when the process of leptogenesis is completed 
at temperatures $T>10^{12}$ GeV for which charged lepton Yukawa processes
are much slower than the processes involving $\widetilde N$ and than
the expansion rate of the Universe. 

However, as we have seen, soft leptogenesis is only effective enough 
for relatively light right-handed sneutrino masses 
$M\lesssim 10^{9}$ GeV. Therefore 
in the relevant temperature window around
$T\sim M$ processes mediated by the $\tau$ and the $\mu$ Yukawa couplings
become faster. As a consequence, the lepton states produced in the 
$\widetilde N$ (and $N$) decay  lose  their coherent between two subsequent
$L$-violating interactions. So before they can re-scatter in reactions
involving $\widetilde N$ and  $N$  they are projected onto the flavour
basis.  In this case, the decay rates and scattering processes involving 
the different flavours $l_k$, $\tilde l_k$ 
and anti-flavours $\bar l_k$, $\tilde l_k^*$ have to be considered 
separately and we need to consider the BE for 
the single lepton-flavour asymmetries.  

To account for this effect we need to define the CP flavour asymmetries
\begin{equation}
\label{epik}
\epsilon^k = \frac{\displaystyle \sum_{a_k} \gamma(\widetilde{N}_\pm \rightarrow a_k)
- \gamma(\widetilde{N}_\pm \rightarrow \bar{a}_k)}
{\displaystyle \sum_{a_k} \gamma(\widetilde{N}_\pm \rightarrow a_k)
+ \gamma(\widetilde{N}_\pm \rightarrow \bar{a}_k)} \ , 
\end{equation}
Taking into account that the Yukawa couplings can be chosen to be real
the flavoured decay rates verify (neglecting the zero temperature
masses)   
\begin{eqnarray}
\gamma(\widetilde{N}_\pm \rightarrow  {a_k})
&=& K^0_k\, \sum_k \gamma(\widetilde{N}_\pm \rightarrow {a_k}) \nonumber \\
\gamma(\widetilde{N}_\pm \rightarrow \bar{a}_k)
&=& K^0_k\, \sum_k \gamma (\widetilde{N}_\pm \rightarrow \bar{a}_k) 
\end{eqnarray}
with projections $K^0_k$
\begin{equation} 
K^0_k= \frac{|Y_{1k}|^2}{{\displaystyle \sum_k}|Y_{1k}|^2}
\end{equation}
so 
\begin{equation}
\epsilon^k(T)\,\equiv \,\bar\epsilon^k\, \Delta_{BF}(T)\,=\,
K^0_k\, 
\bar\epsilon\, \Delta_{BF}(T)\,=\, K^0_k 
\, \epsilon (T) 
\label{eq:barepj}
\end{equation}
with $\epsilon (T)$ given in Eq.~(\ref{eq:asymunf}).
We notice that because of the assumption of alignment between the
soft supersymmetry-breaking $A$ terms and the corresponding neutrino Yukawa 
couplings, the only source of CP violation is a flavour independent phase.
Therefore,  unlike in the case of see-saw leptogenesis induced by 
$N$ decay~\cite{flavour1,flavour2}, in this ``minimal'' 
soft leptogenesis scenario, 
it is not  possible to have non-zero flavour asymmetries with a vanishing 
total  CP asymmetry. 

In Ref.~\cite{flavour1} the relevant equations including flavour
effects associated to the charged--lepton Yukawas were derived
in the density operator approach. One can define a density matrix
for the difference of lepton and antileptons such that 
$\rho_{kk}=Y_{L_k}$. 
As discussed in Ref.~\cite{flavour1,flavour2}
as long as we are in the regime in which a given set of the charged--lepton
Yukawa interactions are out of equilibrium, one can restrict 
the general equation for the matrix density $\rho$ to a subset of
equations for the flavour diagonal directions $\rho_{kk}$. 
In the transition regimes in which a given Yukawa interaction is
approaching equilibrium the off-diagonal entries of the density 
matrix cannot be neglected~\cite{flavour2,riottosc}. However,
as we will see below, for the case of soft leptogenesis, this
is never the case. 

There are additional flavour effects associated to the neutrino Yukawa 
couplings as discussed in Ref.\cite{PU} such as those arising from
processes mediated by  $N_2$ and $\widetilde{N}_2$ (and $N_3$ and
$\widetilde{N}_3$). 
These effects are particularly important in see-saw resonant leptogenesis 
in which right-handed neutrinos of different ``generations'' are close 
in mass.  
In leptogenesis with strong hierarchy among the  masses of the different 
generations of right-handed neutrinos/sneutrinos (as we are assuming
here) one can neglect the neutrino Yukawa couplings in
most of the parameter space, because  the charged--lepton Yukawa rates 
are faster at the temperatures when the asymmetry is produced. 
In what follows we will work under this assumption and neglect flavour
effects associated to the neutrino Yukawas. 

In  writing the BE relevant in the regime in which flavours
have to be considered, it is most appropriate  to follow the evolution of
$Y_{\Delta_{k}}$ where $\Delta_{k}
=\frac{B}{3}-Y_{L_{k_{f}}}-Y_{L_{k_{s}}}\equiv\frac{B}{3}
-Y_{L^k_{\rm tot}}$.
This is so because  $\Delta_{k}$ is conserved by sphalerons and by 
other MSSM interactions. 
In particular, notice that the MSSM processes enforce the equality
of fermionic and scalar lepton asymmetries of the same flavour. Hence, we can
write down the flavoured BE for  $Y_{\Delta_{k}}$

\begin{eqnarray}
sHz\frac{dY_{\Delta_{k}}}{dz} 
& = & -\left\{\epsilon^{k}(T)
\left(\frac{Y_{\tilde{N}_{\mbox{tot}}}}{Y_{\tilde{N}}^{eq}}-2\right) 
\gamma_{\tilde{N}} -\sum_{j}A_{kj}\frac{Y_{\Delta_{j}}} 
{2Y_{c}^{eq}} \, 
\gamma_{\tilde{N}}^{(k)}\right.
\nonumber \\
 &  & -\sum_{j}A_{kj}
\frac{Y_{\Delta_{j}}}{2Y_{c}^{eq}}
\left(\frac{Y_{\tilde{N}_{\mbox{tot}}}}{Y_{\tilde{N}}^{eq}}
\gamma_{t}^{(5)k}+2\gamma_{t}^{(6)k}+2\gamma_{t}^{(7)k}
+\frac{Y_{N}}{Y_{N}^{eq}}\gamma_{t}^{(3)k}+2\gamma_{t}^{(4)k}
\right.\nonumber \\
 &  & \left.\frac{1}{2}\gamma_{N}^{k}+\gamma_{\tilde{N}}^{(2)k}
+\frac{1}{2}\frac{Y_{\tilde{N}_{\mbox{tot}}}}{Y_{\tilde{N}}^{eq}}
\gamma_{t}^{(8)k}+2\gamma_{t}^{(9)k}+2\frac{Y_{N}}{Y_{N}^{eq}}
\gamma_{t}^{(0)k}+2\gamma_{t}^{(1)k}+2\gamma_{t}^{(2)k}
\right)\nonumber \\
 &  & \left.-\sum_{j}A_{kj}\frac{Y_{\Delta_{j}}}
{2Y_{c}^{eq}}\left(2+\frac{1}{2}\frac{Y_{\tilde{N}_{\mbox{tot}}}}
{Y_{\tilde{N}}^{eq}}\right)\gamma_{22}^{k} 
\right\},
\label{eq:BE_L_tot_flavor}
\end{eqnarray}
while the BE for $Y_{N}$ and $Y_{\tilde{N}_{\mbox{total}}}$ remain
the same. 

In Eq.~(\ref{eq:BE_L_tot_flavor}) we have defined the flavoured
thermal widths
\begin{eqnarray}
\gamma_{\tilde{N}}^{k}= K^0_{k}\,\gamma_{\tilde{N}},\\
\gamma_{t}^{(l)k}= K^0_k \,\gamma_{t}^{(l)}
\end{eqnarray}
 
The value of $A_{\alpha\beta}$ will depend on which processes
are in thermal equilibrium when leptogenesis is taking place. For
$T<(1+\tan^{2}\beta)\times10^{9}\mbox{GeV}$
 where the processes
mediated by all the three charged lepton $(e,\mu,\tau)$ Yukawa couplings
are in equilibrium i.e. they are faster than the processes involving
$\tilde{N}_{\pm}$, we have~\cite{antusch}
\begin{equation}
A=\left(\begin{array}{ccc}
-\frac{93}{110} & \frac{6}{55} & \frac{6}{55}\\
\frac{3}{40} & -\frac{19}{30} & \frac{1}{30}\\
\frac{3}{40} & \frac{1}{30} & -\frac{19}{30}
\end{array}\right).
\label{eq:Athree}
\end{equation}
For $(1+\tan^{2}\beta)\times10^{9}\mbox{GeV}<T<(1+\tan^{2}\beta)\times10^{12}\mbox{GeV}$
where only flavours $e+\mu$ and $\tau$ are distinguishable, we have 
\begin{equation}
A=\left(\begin{array}{cc}
-\frac{541}{761} & \frac{152}{761}\\
\frac{46}{761} & -\frac{494}{761}\end{array}
\right).
\label{eq:Atwo}
\end{equation}
For $T>(1+\tan^{2}\beta)\times10^{12}\mbox{GeV}$ when all the flavours
are indistinguishable i.e. the charged lepton Yukawa processes are
much slower than the processes involving $\tilde{N}_{\pm}$, we recover
the unflavoured case where $A=-1.$

From the results in Sec.~\ref{sec:unfsoft} we see that 
in the relevant temperature window around
$T\sim M$ and for $1\leq \tan\beta\leq 30$ we are we are always in
the regime when $T<(1+\tan^{2}\beta)\times10^{9}\mbox{GeV}$ thus we need
to consider flavour effects associated to the three lepton flavours
separately with $A$ given in Eq.~(\ref{eq:Athree})

\section{Results}
\label{sec:results}
We parametrize the asymmetry generated by the decay of the 
singlet sneutrino states in a given flavour as 
\begin{equation}
Y_{\Delta_j}(z\rightarrow \infty)=-2 \eta_j\, \bar\epsilon^j  \, 
Y^{\rm eq}_{\tilde N}(T>>M) 
\label{eq:yb-lfla}
\end{equation}
where $\bar\epsilon^j$ is defined  in Eq.(\ref{eq:barepj}).
Thus the final total asymmetry can be written as Eq.~(\ref{eq:yb-l})
where now 
\begin{equation}
\eta_{fla}=\sum_j \eta_j\, K^0_j 
\end{equation}

\FIGURE[t]{
\includegraphics[width=0.6\textwidth]{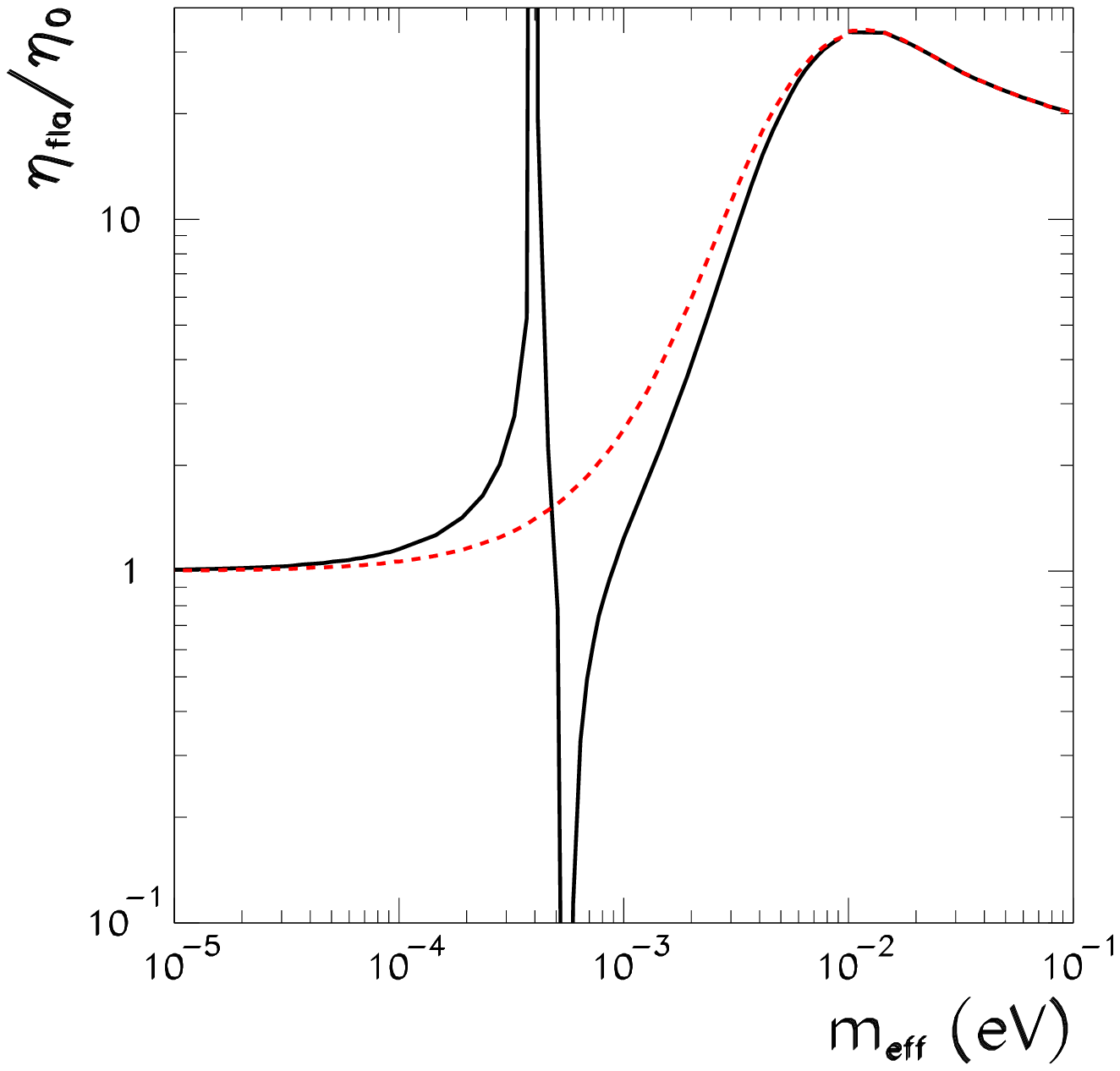}
\caption{Efficiency factor $|\eta/\eta_0 |$ as a function of $m_{eff}$ 
for $M=10^{7}$ GeV and $\tan\beta=30$. 
The two curves
correspond to  vanishing initial $\tilde N$ abundance 
(solid black curve) and thermal initial $\tilde N$ abundance, 
(dashed red curve). 
\label{fig:etafla}}}

In Fig.~\ref{fig:etafla} we plot $|\eta_{fla}/\eta_0|$ 
as a function of $m_{eff}$ for $M=10^7$ GeV 
and for $K^0_1=K^0_2=K^0_3=\frac{1}{3}$.
We label $\eta_0$  the corresponding
efficiency factor without considering flavour effects. 
As seen in the figure for these values of the flavour projections, $K^0_i$, 
and large $m_{eff}$ (large washout region), flavour effects can 
make leptogenesis more efficient by up to a factor of the order 
$30$. On the contrary flavour effects play no role for 
small $m_{eff}$ (small washout). 
This can be easily understood by adding the equations for the three 
flavour asymmetries, 
Eq.~(\ref{eq:BE_L_tot_flavor}). 
We get an equation which can be written as:
\begin{equation}
sHz\frac{dY_{B-L}}{dz} 
=  -\left\{\epsilon(T)
\left(\frac{Y_{\tilde{N}_{\mbox{tot}}}}{Y_{\tilde{N}}^{eq}}-2\right) 
\gamma_{\tilde{N}} -\sum_{kj}\,A_{kj}\, K^0_k\, 
\frac{Y_{\Delta_{j}}}{2Y_{c}^{eq}} \,  W \right\}
\label{eq:Be_fla_sum}
\end{equation} 
where we have defined the washout term 
\begin{eqnarray}
W&=& \gamma_{\tilde{N}}
+\frac{Y_{\tilde{N}_{\mbox{tot}}}}{Y_{\tilde{N}}^{eq}}
\gamma_{t}^{(5)}+2\gamma_{t}^{(6)}+2\gamma_{t}^{(7)}
+\frac{Y_{N}}{Y_{N}^{eq}}\gamma_{t}^{(3)}+2\gamma_{t}^{(4)}
+\frac{1}{2}\gamma_{N}+\gamma_{\tilde{N}}^{(2)} \nonumber 
\\ 
&&
+\frac{1}{2}\frac{Y_{\tilde{N}_{\mbox{tot}}}}{Y_{\tilde{N}}^{eq}}
\gamma_{t}^{(8)}+2\gamma_{t}^{(9)}+2\frac{Y_{N}}{Y_{N}^{eq}}
\gamma_{t}^{(0)}+2\gamma_{t}^{(1)}+2\gamma_{t}^{(2)}
+\left(2+\frac{1}{2}\frac{Y_{\tilde{N}_{\mbox{tot}}}}
{Y_{\tilde{N}}^{eq}}\right)\gamma_{22} 
\end{eqnarray}
which can be directly compared with the unflavoured equation 
Eq.~(\ref{eq:BE_L_tot}). We see that if we define 
$P_j =Y_{\Delta_{j}}/Y_{B-L}$, Eq.~(\ref{eq:Be_fla_sum}) is equivalent
to Eq.~(\ref{eq:BE_L_tot}) with 
\begin{equation}
W \rightarrow - W \, \times\, \sum_{ij} A_{ij} K^0_j P_i
\end{equation}
Thus flavour effects are unimportant
when the $W$ term in Eq.~(\ref{eq:Be_fla_sum}) 
is much smaller than the source term which happens 
when $m_{eff}$ is small enough (small washout regime).

We have verified that 
the equally distributed flavour composition 
$K^0_1=K^0_2=K^0_3=1/3$ (so all flavour are in the same washout regime) 
gives an almost maximum flavour effect for $m_{eff}\lesssim 10^{-2}$ eV. 
Conversely for $m_{eff}\gtrsim 10^{-2}$ eV values, flavour effects 
lead to  larger  $B-L$ for more  asymmetric flavour compositions.
In this case, the ``optimum'' flavour projection strongly depends
on the value of $m_{eff}$. 

\FIGURE[t]{
\includegraphics[width=\textwidth]{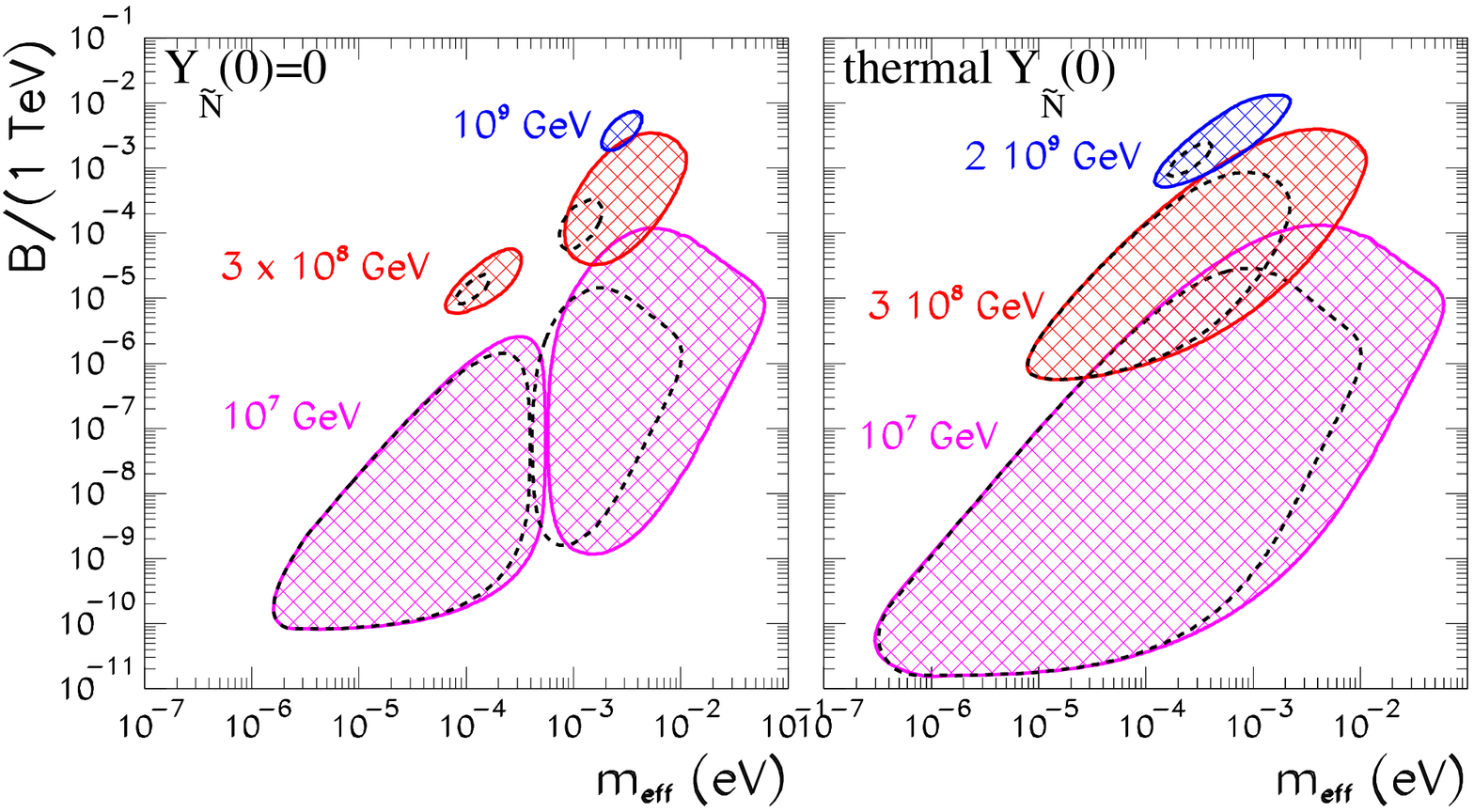}
\caption{
$B, m_{eff}$ regions in which successful soft leptogenesis can
be achieved when flavour effects are included with $K^0_1=K^0_2=K^0_3=1/3$
We take $|{\rm Im} A|=10^3$ GeV and $\tan\beta=30$ and different
values of $M$  as labeled in the figure.
The dashed contours are the corresponding ones when flavour effects
are not included. 
The two panels correspond to  vanishing initial $\tilde N$ abundance 
(left) and thermal initial $\tilde N$ abundance, 
(right). 
\label{fig:Bmefffla}}}

We plot in  Fig. \ref{fig:Bmefffla} the range of parameters
$B$ and $m_{eff}$ for which enough asymmetry is
generated, $Y_B\geq 8.54\times 10^{-11}$ for the 
the equally distributed flavour composition 
$K^0_1=K^0_2=K^0_3=1/3$.
We show the ranges for several values of $M$  and 
for the characteristic value of  $|{\rm Im} A|=1$ TeV.
The dashed contours are the corresponding ones when flavour effects
are not included. 
The figure illustrates to what extent flavour effects 
can affect the ranges of $B$ and $M$ for which successful soft 
leptogenesis can be achieved. This is more quantitatively displayed in  
Fig.~\ref{fig:bmmax} where we plot 
the  asymmetry that can 
be achieved for a give value of $B$ (or $M$) maximized with respect
to $m_{eff}$ and $M$ (or $B$) when flavour effects are included
(for $K^0_1=K^0_2=K^0_3=1/3$) compared to the corresponding one when
they are neglected. From the figure we read that successful 
soft-leptogenesis with (without) flavour effects considered
requires 
$B\leq 8\times 10^{-3}$ ($3\times 10^{-4}$) 
TeV and $M\leq\times 10^9$ ($4\times 10^8$)  GeV
for vanishing initial $\tilde N$ abundance and 
$B\leq 1.5\times 10^{-2}$ ($3\times 10^{-3}$) 
TeV and $M\leq 3\times 10^9$ ($2\times 10^9$)  GeV 
for thermal initial $\tilde N$ abundance.

\FIGURE[t]{
\includegraphics[width=0.8\textwidth]{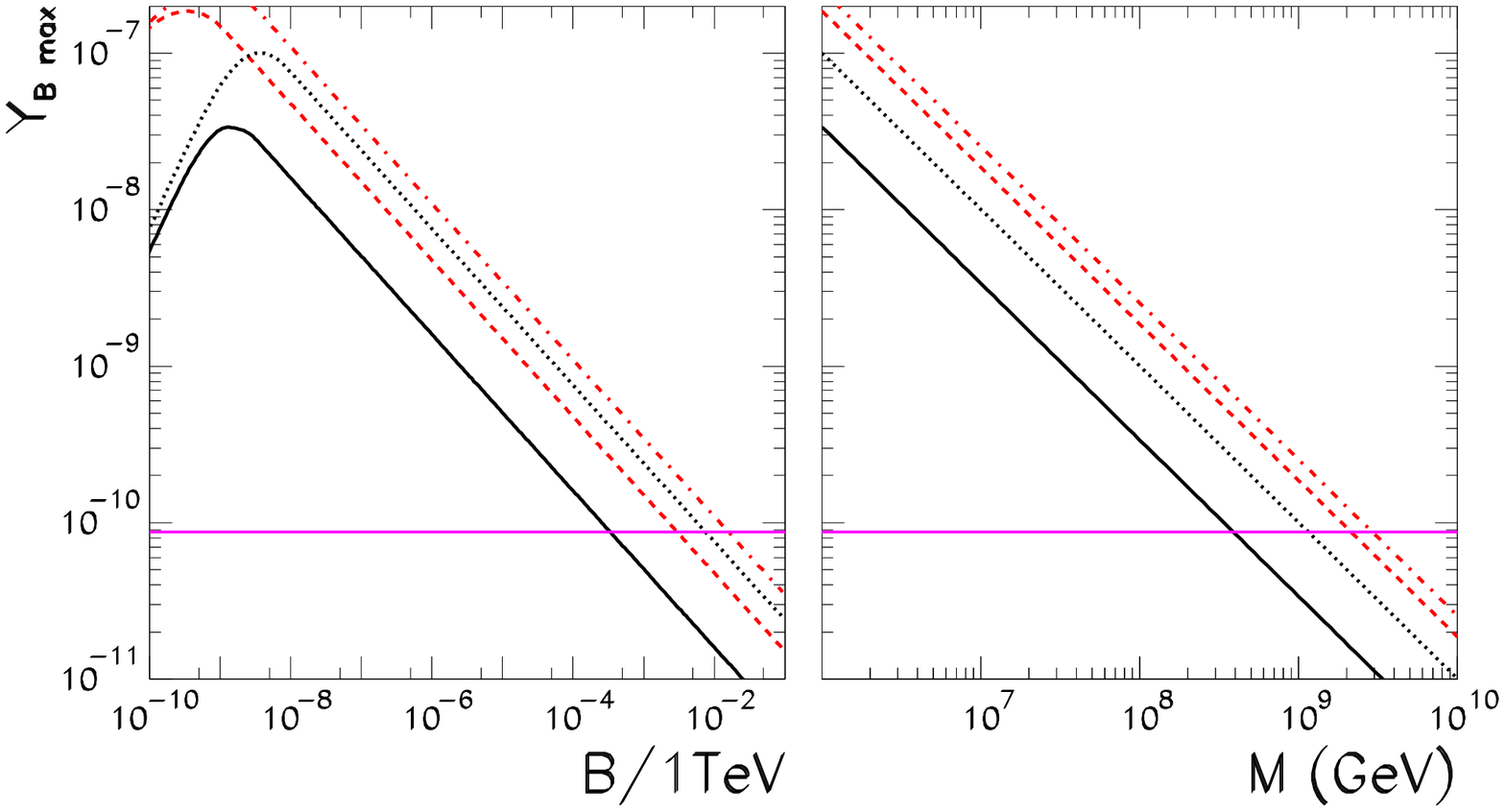}
\caption{Maximum baryon asymmetry 
be achieved as a function of $B$ (left) and $M$ (right).
The solid (dashed) lines are for no flavour
effects and vanishing (thermal) initial $\tilde N$ abundance.
The dotted (dash-dotted) lines are the corresponding asymmetries
after including flavour effects with $K^0_1=K^0_2=K^0_3=1/3$.
The horizontal line correspond to the $1\sigma$  
WMAP measurements  in the $\Lambda$CDM model Eq.~(\ref{eq:wmapeta}). 
\label{fig:bmmax}}}
 
In summary,  in this work we have studied the impact of flavour in 
soft leptogenesis. We have quantified to what extent flavour effects,
which must be accounted for in the relevant sneutrino mass range, can  
affect the region of parameters in which successful 
leptogenesis induced by CP violation in the right-handed sneutrino 
mixing is possible. 
We find that for decays which occur 
in the intermediate to strong washout regimes for all flavours, 
the produced total $B-L$ asymmetry can be up to a factor ${\cal O}(30)$
larger than the one predicted with flavour effects being neglected.
This enhancement, permits slightly larger values of the required 
lepton violating  soft bilinear term  $B$. 

\acknowledgments 
We thank Y. Nir for careful reading of the manuscript.
This work is supported by National Science Foundation
grant PHY-0354776 and  by Spanish Grants FPA-2004-00996
and  FPA2006-28443-E.

\end{document}